# Above 400 K Robust Perpendicular Ferromagnetic Phase in a Topological Insulator

Chi Tang[1]†, Cui-Zu Chang[2,3]†, Gejian Zhao[4], Yawen Liu[1], Zilong Jiang[1], Chao-Xing Liu[3], Martha R. McCartney[4], David J. Smith[4], Tingyong Chen[4], Jagadeesh S. Moodera[2,5], and Jing Shi[1]*

Use superscript numbers to link affiliations, and symbols *†‡ for author notes.

[1]Department of Physics & Astronomy, University of California, Riverside, CA 92521, USA.

[2]Francis Bitter Magnet Laboratory, Massachusetts Institute of Technology, Cambridge, MA 02139, USA.

[3]Department of Physics, The Pennsylvania State University, University Park, PA 16802, USA.

[4]Department of Physics, Arizona State University, Tempe, AZ 85287, USA.

[5]Department of Physics, Massachusetts Institute of Technology, Cambridge, MA 02139, USA

*Correspondence to: jing.shi@ucr.edu.

†These are co-first authors.

**Abstract:** The quantum anomalous Hall effect (QAHE) that emerges under broken time-reversal symmetry in topological insulators (TI) exhibits many fascinating physical properties for potential applications in nano-electronics and spintronics. However, in transition-metal doped TI, the only experimentally demonstrated QAHE system to date, the effect is lost at practically relevant temperatures. This constraint is imposed by the relatively low Curie temperature ($T_c$) and inherent spin disorder associated with the random magnetic dopants. Here we demonstrate drastically enhanced $T_c$ by exchange coupling TI to $Tm_3Fe_5O_{12}$, a high-$T_c$ magnetic insulator with perpendicular magnetic anisotropy. Signatures that the TI surface states acquire robust ferromagnetism are revealed by distinct squared anomalous Hall hysteresis loops at 400 K. Point-contact Andreev reflection spectroscopy confirms that the TI surface is indeed spin-polarized. The greatly enhanced $T_c$, absence of spin disorder, and perpendicular anisotropy are all essential to the occurrence of the QAHE at high temperatures.

**One Sentence Summary:** We unequivocally demonstrate proximity-induced ferromagnetism with perpendicular anisotropy persisting well above 400 K in the topological Dirac surface states.

# INTRODUCTION

The discovery of the quantum anomalous Hall effect (QAHE) in magnetically-doped topological insulators (TI) has spurred a great deal of excitement in the study of the topological state of matter *(1-3)*. Unlike the well-known quantum Hall effect or regular anomalous Hall effect (AHE), the remarkable QAHE state is characterized by the quantized Hall conductance $\pm e^2/h$ and dissipation-less chiral edge-current transport in the absence of any external magnetic field. Furthermore, in the QAHE state the chiral edge-charge current simultaneously carries a large and electrically tunable spin polarization due to strong spin-orbit coupling in TI *(4)*. These attractive properties make QAHE materials strong candidates for next generation spintronic applications at the nanoscale.

An essential pre-requisite for the QAHE to occur is an exchange interaction that breaks the time-reversal symmetry in TI *(1, 5)*. In magnetically-doped TI, exchange interaction originates from the long-range ferromagnetic order of magnetic moments carried by transition metal dopants such as Cr, V and Mn *(2, 3, 6)*. The spontaneous magnetic order leads to an exchange gap at the Dirac point of TI, which consequently gives rise to the QAHE. Hence, the ultimate upper temperature limit of the QAHE is dictated by the Curie temperature ($T_c$) of the magnetic TI. In the optimal Cr- or V-doped TI samples that exhibit QAHE, $T_c$ can be as high as 30 K, but the QAHE only occurs at much lower temperature, i.e. at or well below 2 K *(2, 3, 7-9)*. While the origin of this large discrepancy is an interesting topic for further investigation, it is imperative to drastically raise the $T_c$ and therefore greatly increase the exchange-induced gap, to ultimately realize QAHE at elevated temperatures.

One possible route to higher $T_c$ is to continue optimizing the transition-metal doping approach, similar to previous efforts in dilute magnetic semiconductor research *(10)*. However, raising $T_c$ requires increasing the doping level, which inevitably lowers sample quality with the presence of extensive disorder, or even destroys the nontrivial surface states due to weakened spin-orbit coupling *(11)*. An alternative route is to leverage the proximity effect to couple the TI surface states directly to a high-$T_c$ magnetic insulator without introducing spin disorder due to randomly doped magnetic moments to either bulk or surface states. The latter approach has an obvious advantage of independent optimization of both electronic and magnetic properties *(12)*, which is impossible in the first approach. Recent progress in this direction, including TI/EuS *(13-16)*, $Y_3Fe_5O_{12}$(YIG)/TI *(12, 17-19)* and CrSb/TI *(20)* heterostructures, indicates its feasibility. However, while induced exchange interaction is necessary for QAHE, the in-plane magnetization in a system such as YIG/TI does not open an exchange gap in the TI surface electronic states unless reflection symmetry is broken *(21)*. In this *Report*, we demonstrate induced ferromagnetism in TI with perpendicular magnetic anisotropy, as manifested by sharp and squared AHE hysteresis loops, using a high-$T_c$ (~ 560 K) ferrimagnetic insulator, namely, $Tm_3Fe_5O_{12}$ or TIG. In addition, the perpendicular ferromagnetic phase in TI persists well above 400 K, which is more than one order of magnitude higher than the $T_c$ of the optimal magnetically-doped TI exhibiting QAHE. These unique features entail a much larger exchange gap in TI without requiring any external magnetic field.

# RESULTS

TIG/TI heterostructures

Similar to the YIG films in previous studies *(12, 17)*, TIG is a rare-earth garnet with ferrimagnetism originating from the anti-ferromagnetically coupled iron magnetic moments via super-exchange interaction. To ensure high heterostructure quality, we first prepare atomically flat TIG surfaces while engineering the needed perpendicular magnetic anisotropy. This is accomplished by epitaxial growth of TIG films on (111)-oriented substituted gadolinium gallium garnet (SGGG) substrates with pulsed laser deposition *(22, 23)*. Due to the negative magnetostriction constant of TIG, the interface tensile strain exerted by SGGG produces perpendicular magnetic anisotropy *(22)*. The root-mean-square surface roughness in 10-nm-thick TIG films can be as low as ~ 1.2 Å. Meanwhile, the resulting perpendicular anisotropy is sufficiently strong to drive the magnetization in the direction normal to the film plane without any external magnetic field. TIG/TI heterostructures are subsequently fabricated by growing a five quintuple-layer (QL) thick $(Bi_xSb_{1-x})_2Te_3$ TI film atop TIG using molecular beam epitaxy (MBE) *(12, 17, 24)*. These TI films show excellent structural quality, as confirmed by *in situ* reflection high-energy electron diffraction (RHEED) patterns (see Fig. S1A). *Ex situ* high-resolution transmission electron microscopy (HRTEM) does not show any visible interfacial defects or additional phases (Fig. 1C).

The chemical potential of the TI films can be controlled by varying the Bi:Sb ratio, as demonstrated previously *(25)*. To study the effect of carrier type on interface ferromagnetism, we deliberately choose two ratios, $x=0.20$ and $x=0.30$, in order to place the chemical potential just below and above the Dirac point, respectively. Both films behave as bulk semiconductors with the chemical potential inside the bulk gap, as indicated by the typical temperature dependence of the longitudinal sheet resistance ($R_{xx}$) *(25)*, as shown in the insets of Figs. 1D and 1E. Figs. 1D and 1E show Hall traces of these two samples measured at 400 K. Besides the clear hysteresis loops, which are further discussed below, the linear ordinary Hall background indicates opposite carrier types in these two samples, with hole carrier density ($n_{2D}$) of $3.3\times10^{13}$ cm$^{-2}$ for $x=0.20$ and electron $n_{2D}$ of $3.5\times10^{13}$ cm$^{-2}$ for $x=0.30$ at $T=400$ K. In consideration of the carrier densities below $1\times10^{13}$ cm$^{-2}$ at $T=2$ K and the thermally excited additional carriers at higher temperatures, the chemical potential is indeed located on either side of the Dirac point of the surface states but still within the bulk gap.

Anomalous Hall Effect

A squared Hall hysteresis loop, which resembles the TIG magnetic hysteresis loop taken with an out-of-plane magnetic field (Fig. S1B), is superimposed on the linear background. In a normal ferro- or ferri-magnetic conductor, this type of loop would be the standard anomalous Hall signal characteristic of the ferromagnetism of the conductor *(26)*. However, the ferrimagnet here is an insulator, and the Hall response can only come from the TI film. The hysteresis must be acquired by the bottom surface of the TI which is in direct contact with the TIG. More representative Hall loops above room temperature are presented in Fig. 2 (low-temperature data are shown in Figs. S2 and S3). The slight enhancement of the linear Hall slope at lower temperature displayed in Figs. 2A and 2C is probably caused by a shift of the chemical potential or narrowing of the Fermi-Dirac distribution function which changes the carrier density. After the linear ordinary Hall background is subtracted, the corresponding AHE hysteretic loops are displayed in Figs. 2B and 2D, respectively. The remarkable presence of sharp and squared loops indicates a strongly preferred perpendicular magnetization direction. Clearly, such strong perpendicular anisotropy exists throughout the entire temperature range up to 400 K (also shown in Fig. S6), the highest temperature in our measurements. It is apparent from the high temperature trend that the hysteresis extends well above 400 K, which is more than an order of magnitude higher than the record achieved in modulation-doped TI films exhibiting the QAHE *(2, 3, 7-9)*. As the temperature decreases, the magnitude of the AHE resistance steadily increases but the sign remains unchanged, as shown in Fig. S4, which is accompanied by an increase in $R_{xx}$. This temperature dependence is presumably due to the reduction of bulk carriers when the temperature is lowered. It is interesting to note that the coercive fields of the two samples are different. This difference was also observed in TIG/Pt samples and attributed to the coercive field variations due to the local strain variation at the SGGG/TIG interface *(22)*. The increased coercive field at lower temperatures results from the magnetic anisotropy increase in TIG, likely due to an enhanced magneto-elastic coefficient at low temperatures *(27)*. Importantly, the AHE loops have the same sign in these two samples despite the different carrier types. This feature immediately excludes the possibility of any Lorentz force-induced responses arising either from TIG stray fields or contributions from two types of carriers in TI.

Spin Polarization Determined by Andreev Reflection Spectroscopy

A straightforward mechanism for the squared AHE hysteresis loops is induced ferromagnetism due to proximity coupling with TIG. Alternatively, since TI has strong spin-orbit coupling, the spin Hall effect could in principle give rise to similar AHE loops *(22)*, and this scenario does not require any induced ferromagnetism in TI. However, it is only the first mechanism that produces the exchange gap relevant to the QAHE. Magnetoresistance measurements can shed some light on the relative importance of the spin Hall effect. We have performed those measurements with both rotating and sweeping magnetic fields and the results point towards the first mechanism (Figs. S5 and S7). However, unambiguously confirming the exact mechanism with magneto-transport remains a major challenge. To directly verify the existence of induced ferromagnetism in TI surface, we employ point-contact Andreev reflection spectroscopy (PCAR), a well-established probe for quantitatively measuring spin polarization in ferromagnetic materials *(28, 29)*. Since the current only flows in the TI part of the heterostructures, it exclusively detects spin polarization in the TI. To perform the measurements, the heterostructure sample was first cleaved. The superconducting tip then locates and approaches the interface from the cleaved side. Because of a limitation on tip size, the tip-sample contact is across the entire TI thickness, as illustrated by the inset of Fig. 3. Consequently, both the top and bottom surfaces of the TI layer contribute to the Andreev spectrum. Due to the spin-momentum locking, electrons on both TI surfaces have opposite spins in the film plane; thus, zero spin polarization is expected if both surfaces contribute equally. This is indeed observed in the 20 QL sapphire/TI reference sample, as shown by the dark blue curve in Fig. 3. An analysis using the modified BTK model *(30)* shows spin polarization ($P$) as small as 1%, which suggests slight asymmetry between two surfaces. In contrast, in TIG/$(Bi_{0.20}Sb_{0.80})_2Te_3$(20 QL), $P$ is found to be over 30% with a large interfacial factor or Z-factor ($Z = 0.4$), much greater than that in the reference sample. As shown in Fig. 3, the Andreev peak of the TIG/TI sample is much lower than that of the sapphire/TI, also indicating higher $P$. In PCAR, $P$ is often significantly reduced by having a finite interfacial scattering Z-factor. The observed $P$ of 30% at a large Z-factor of 0.4 is therefore underestimated. Due to the difficulty in making transparent side contacts, we are unable to realize low Z-factor contacts. However, from several relatively large Z-factor contacts, we can extrapolate to an intrinsic $P$ value of 70% at $Z = 0$ (see Fig. S8), unequivocally demonstrating proximity-induced ferromagnetism in TI surface electronic states at the measurement temperature. Note that the AHE signal monotonically decreases as the temperature is increased (Fig. S4), which is consistent with the decreasing trend of both induced magnetization and longitudinal resistivity. More experimental investigations are underway to directly measure the spin polarization at high temperatures.

In conclusion, we have demonstrated robust above 400 K AHE which is dominated by proximity-induced ferromagnetism with strong perpendicular magnetic anisotropy in TI surface states of the TIG/TI heterostructures. Similar proximity-induced exchange interaction has been observed in ferromagnet/TI and anti-ferromagnet/TI heterostructures *(12-20)*, but the effect in our heterostructures extends at much higher temperatures. The proximity approach offers a powerful yet flexible method to tailor both electronic and magnetic properties to realize the QAHE much above room temperatures for potential applications. The TIG/TI heterostructure also provides an ideal platform to explore other topological magneto-electric effects *(5, 31)*, such as quantized Faraday and Kerr rotation *(32-34)* and image magnetic monopole *(35)*.

## MATERIALS AND METHODS

Thin TIG films are grown on epi-ready single crystal SGGG (111) substrates via pulsed laser deposition. SGGG and TIG have sufficiently large lattice mismatch so that they are chosen to generate a tensile strain to induce a large perpendicular magnetic anisotropy in TIG due to its negative magnetostriction constant. The base pressure of the deposition chamber is ~ $6\times10^{-7}$ Torr. After going through standard cleaning, the substrates are annealed at ~ 200 ºC for over five hours prior to TIG deposition. The TIG films are then grown at moderate temperature ~ 500 °C by KrF excimer laser pulses of 248 nm in wavelength with power of 170 mJ at a repetition of 1 Hz under 2 mTorr oxygen pressure with 12 wt.% ozone. Rapid thermal annealing process are performed at 850 °C for 5 minutes to magnetize the TIG films. Under this optimal growth condition, the as-grown TIG films exhibit strong perpendicular magnetic anisotropy and ultra-flat surfaces characterized by vibrating sample magnetometry (Fig. S1B), ferromagnetic resonance (19), and atomic force microscopy (19).

To fabricate high-quality TIG/TI heterostructures, TIG (111) films are then transferred to a custom-built ultra-high vacuum MBE system with base pressure better than $5\times10^{-10}$ Torr for TI growth. To ensure good interface quality, *in situ* high temperature annealing (600 °C, 60 mins) is performed before the film growth. The RHEED pattern is taken again to ensure same excellent quality of TIG surface (Fig. S1A inset) after annealing. High-purity Bi (99.999%), Sb (99.9999%), and Te (99.9999%) are evaporated from Knudsen effusion cells. During the growth, the TIG substrate is kept at 230 °C and the growth rate is ~0.2 QL/min. The epitaxial growth is monitored by the *in situ* RHEED pattern. The sharp and streaky diffraction spots indicate a very flat surface and high quality crystalline TI thin film grown on TIG (111) (Fig. S1A). The film is covered with a 5 nm Te protection layer before taken out of the MBE chamber.

The HRTEM micrographs of the TIG/TI heterostructure are recorded using an image-corrected FEI Titan 80-300 operated at 300 kV, with the spherical aberration coefficient set to be ~ +5 microns. The specimen is cross-sectioned for observation by focused-ion-beam milling using an FEI dual-beam Nova 200, with initial milling at 30 kV, and final surface cleaning at 5 kV.

TI films are patterned into Hall bars with width 50 μm and length 300 μm using standard photolithography and inductively coupled plasma etching. Cr (8 nm)/Au (50 nm) is then deposited in the Hall bar contact areas for better contact. Transport measurements from 2 to 400 K are carried out using standard four-terminal dc method in a Quantum Design Physical Property Measurement System. We use Keithley 2400 as a current source meter, Keithley 2182A and Keithley 2000 as voltmeters. The dc source current is kept constant for measurements over a temperature range but adjusted between 5 and 10 μA in different temperature ranges to avoid overheating the sample and maintain a good signal-noise-ratio.

To perform the PCAR measurements, a gold film of ~50 nm is deposited on half of the TI sample to serve as one electrode. A few scratches are made on the sample before deposition so that the gold layer can contact both the top and bottom TI surfaces. The sample is then cleaved to have a fresh cross section and immediately mounted on a home-built Andreev reflection spectroscopy probe, then enclosed into a vacuum jacket which is subsequently pumped to ~$2\times10^{-7}$ Torr, and then filled with helium exchange gas of 0.1 Torr. The probe is cooled down to 4.2 K in a sample tube where liquid helium is introduced through a needle valve, and 1.5 K is realized by pumping the sample tube. The point contacts are established after the temperature has been stabilized. Subsequently, the conductance (*I/V*) and differential conductance (*dI/dV*) are measured simultaneously.

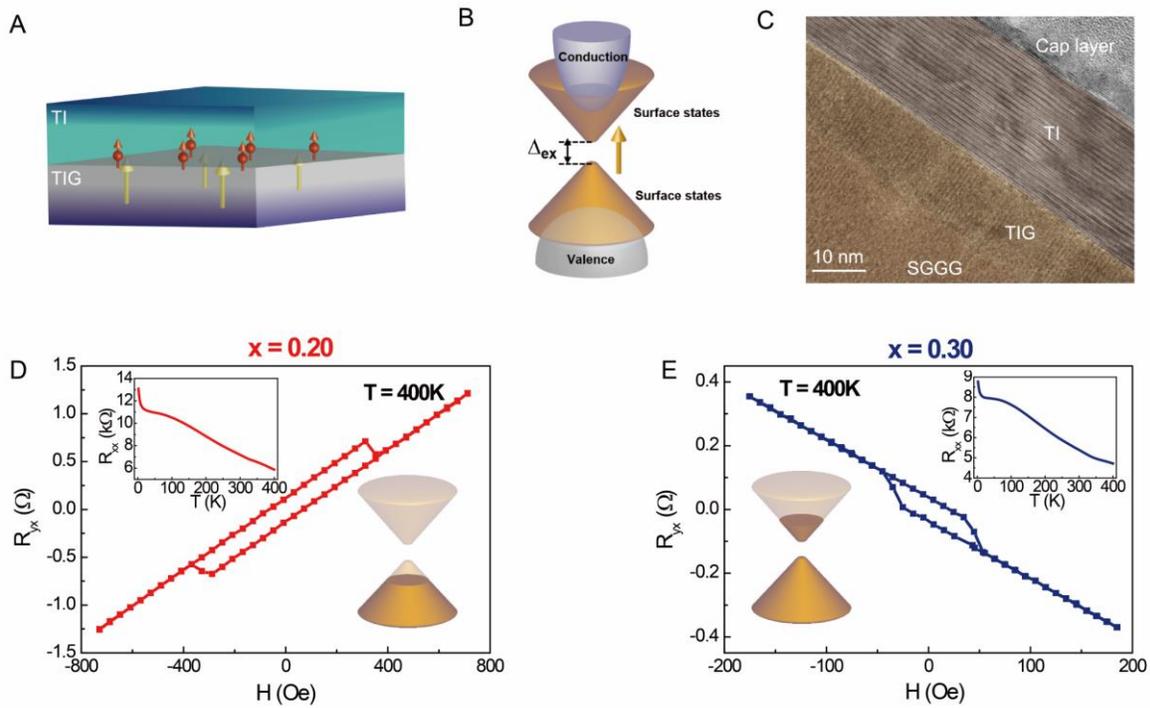

Figure 1. **Proximity-induced ferromagnetism and anomalous Hall effect at 400 K in TIG/TI heterostructure.** A. Schematic drawing of proximity coupling between TI and TIG. B. Exchange gap at the charge neutral point of TI surface states induced by broken time reversal symmetry. C. HRTEM image of a TIG/TI (20 QL) bilayer structure. D & E. Hall traces of TIG/$(Bi_xSb_{1-x})_2Te_3$ (5 QL) for $x = 0.20$ and $x = 0.30$, respectively. The upper insets show the corresponding temperature dependence of $R_{xx}$. The lower insets show schematic drawings of the corresponding chemical potential position.

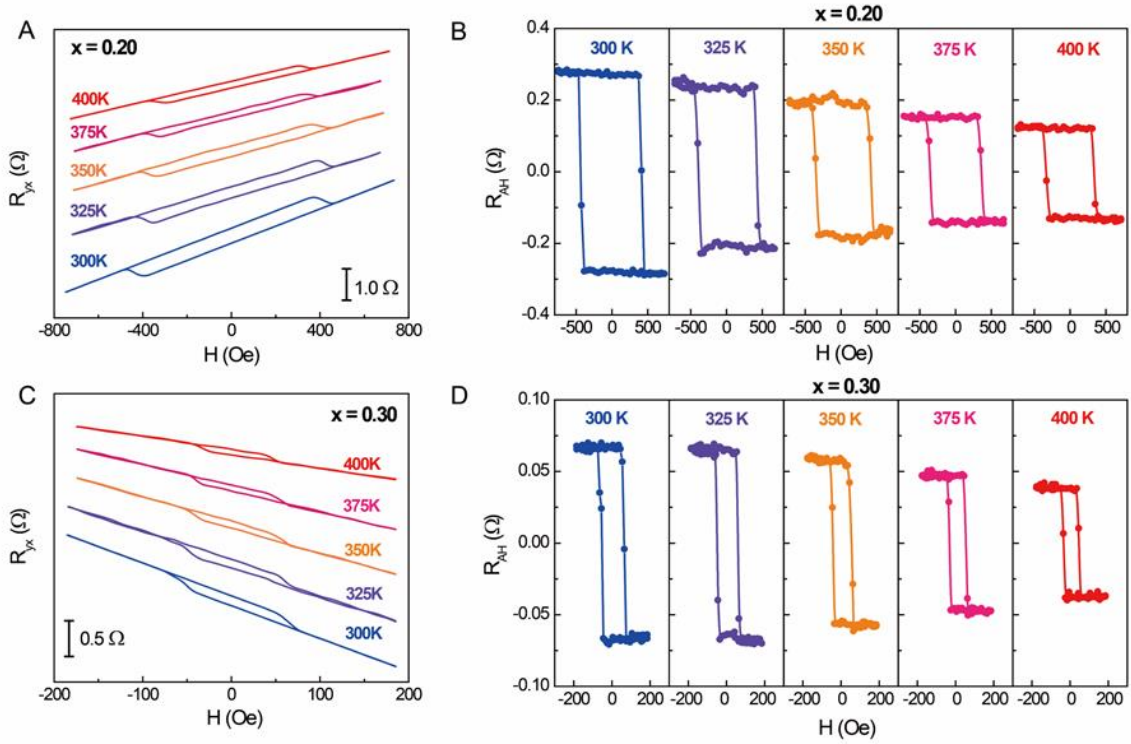

Figure 2. **Temperature dependence of AHE response up to 400 K.** A. Hall resistance of TIG/(Bi$_x$Sb$_{1-x}$)$_2$Te$_3$ (5 QL) for $x = 0.20$ (*p*-type) between 300 and 400 K. B. Temperature dependence of AHE loops in in A after subtracting the linear ordinary Hall background. C. Hall resistance of TIG/(Bi$_x$Sb$_{1-x}$)$_2$Te$_3$ (5 QL) for $x = 0.30$ (*n*-type) between 300 and 400 K. D. Temperature dependence of AHE loops in C after subtracting the linear ordinary Hall background.

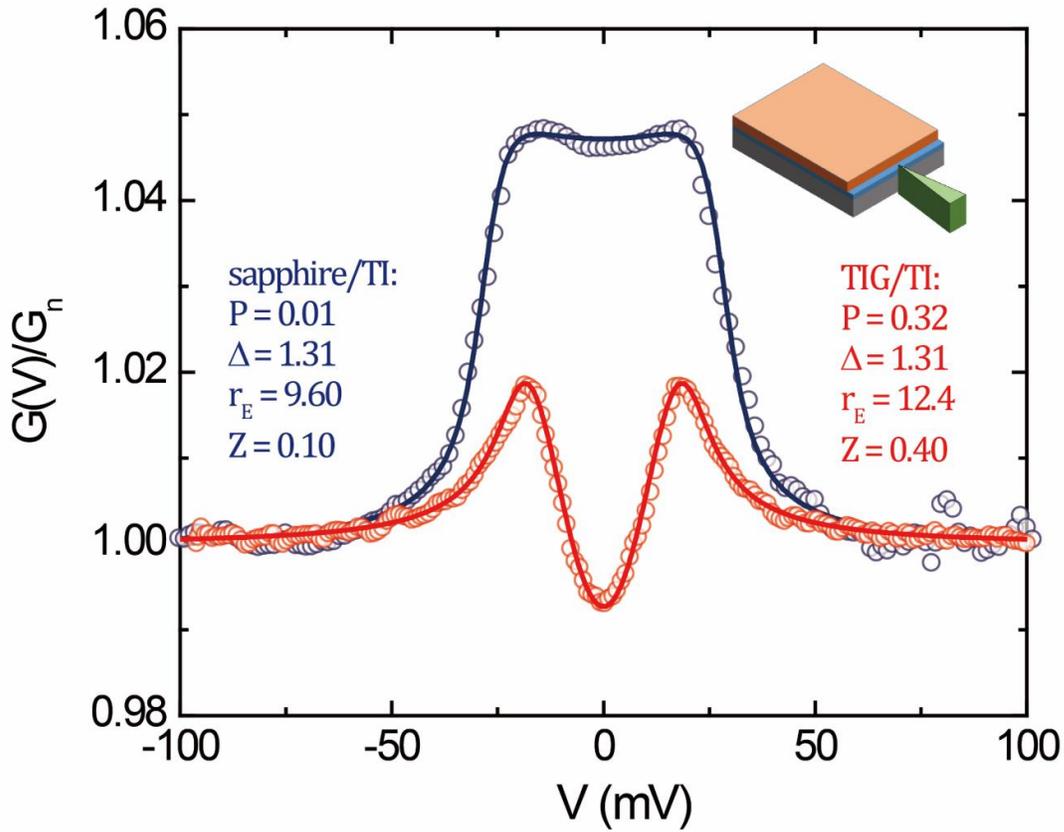

Figure 3. **Representative Andreev reflection spectra for (Bi$_{0.20}$Sb$_{0.80}$)$_2$Te$_3$(20 QL) on TIG and sapphire substrate.** Fitting the normalized differential conductance to the modified BTK model yields *P* of 32% for TI on TIG and near-zero *P* for TI on sapphire. All measurements were taken at 1.5 K. The circles are raw data and solid curves are the best fits. The inset is a schematic drawing of PCAR experiments.

**Acknowledgments:**

The authors thank D. Yan, J. Li, Z. Shi, and N. Samarth for their help with experiments and discussions. **Funding:** The work at UCR and ASU was supported as part of the SHINES, an Energy Frontier Research Center funded by the U.S. Department of Energy, Office of Science, Basic Energy Sciences under Award No. SC0012670. CZC and JSM. acknowledge the support from NSF Grants No. DMR-1207469, No. DMR- 0819762 (MIT MRSEC), ONR Grant No. N00014-13-1- 0301, and the STC Center for Integrated Quantum Materials under NSF Grant No. DMR-1231319. CXL acknowledges the support from ONR Grant No. N00014-15-1-2675. MM and DS acknowledge use of facilities at the John M. Cowley Center for High Resolution Electron Microscopy at ASU. **Competing interests:** The authors declare that they have no competing interests. **Data and materials availability:** All data needed to evaluate the conclusions in the paper are present in the paper and/or the Supplementary Materials. Additional data related to this paper may be requested from the authors.

**Supplementary Materials:**

Figs. S1 to S8